\documentclass[eqsecnum,preprint,showpacs,amsmath,aps,nofootinbib]
  {revtex4}  
\usepackage{epsfig}
\usepackage{amssymb}
\usepackage[dvips]{color}
\usepackage[latin1]{inputenc}
\usepackage{graphicx}

\makeatletter
\renewcommand*{\@fnsymbol}[1]{\ensuremath{\ifcase#1\or *\or \dagger\or
    \ddagger\or 
   \mathsection\or **\or \dagger\dagger
   \or \ddagger\ddagger \else\@ctrerr\fi}}
\makeatother

\begin{document}
\title{Size of the Gribov region in curved spacetime}

\author{Marco de Cesare}
\email[E-mail: ]{marco.decesare5@studenti.unina.it}
\affiliation{Dipartimento di Fisica, Complesso Universitario di
Monte S. Angelo,\\ 
Via Cintia Edificio 6, 80126 Napoli, Italy}

\author{Giampiero Esposito}
\email[E-mail: ]{gesposit@na.infn.it}
\affiliation{Istituto Nazionale di Fisica Nucleare, Sezione di
Napoli, Complesso Universitario di Monte S. Angelo, 
Via Cintia Edificio 6, 80126 Napoli, Italy}

\author{Hossein Ghorbani} 
\email[E-mail: ]{pghorbani@ipm.ir}
\affiliation{School of Particles and Accelerators,\\
Institute for Research in Fundamental Sciences (IPM),\\
P.O. Box 19395-5531, Tehran, Iran}

\date{\today}

\begin{abstract}
Recent work in the literature has argued that the joint effect of
spacetime curvature and the Gribov ambiguity may introduce 
further modifications to the Green functions in the infrared.
This paper focuses on a simple criterion for studying the effect
of spacetime curvature on the size of the Gribov region, improving
the accuracy of the previous analysis. It is shown that, depending
on the sign of the scalar and Riemann curvature, the Gribov horizon
moves inward or, instead, outward with respect to the case of flat
spacetime. This is made clear by two novel inequalities here derived
for the first time. 
\end{abstract}


\maketitle

\section{Introduction}

Ever since Gribov \cite{gribov78} 
and other authors \cite{halp78} discovered the limitations
of quantum Yang--Mills theory in the Coulomb gauge, many efforts
were devoted to studying the issue in a variety of contexts.
In general, in non-perturbative quantum gauge theory, it may
happen that a gauge orbit intersects the surface defined by the
gauge-fixing condition at more than one point. This leads to a
ghost operator having zero-modes, i.e. non-vanishing eigenfunctions
belonging to the zero eigenvalue. The preservation of the
gauge-fixing condition under gauge transformations is then expressed
by a partial differential equation admitting a number of solutions
rather than a unique solution, a property usually studied through
the so-called Gribov pendulum example \cite{gribov78, espo04, mdc13}.
 
In our paper, we study the Gribov problem in curved spacetime, motivated
by the following results:
\vskip 0.3cm
\noindent
(i) The work in \cite{canfo10} has shown that, in a curved background,
a proper gauge fixing cannot be achieved, not even in the Abelian case.
\vskip 0.3cm
\noindent
(ii) When black holes, neutron stars, quarks and hybrid stars, and
cosmological setups are studied, it is important to consider the
dynamics of quantum chromodynamics on a curved background \cite{canfo11}.
\vskip 0.3cm
\noindent
(iii) The coupling to the gravitational field destroys the perturbative
renormalizability of the Yang--Mills field with field strength 
$F_{\; \mu \nu}^{\alpha}$ even in the purely Yang--Mills sector 
\cite{dewi03}. In addition to the familiar term in 
$F_{\alpha \mu \nu}F^{\alpha \mu \nu}$ in the heat-kernel 
$a_{2}$ coefficient, there are now not only the usual terms in
$$
R^{2}, \; R_{\mu \nu}R^{\mu \nu}, 
\; R_{\mu \nu \sigma \tau}R^{\mu \nu \sigma \tau},
$$
but also terms in \cite{dewi03}
$$
\mu^{-2}F_{\alpha \rho \nu ;}^{\; \; \; \; \; \; \; \nu}
F_{\; \; \; \; \; \; ; \sigma}^{\alpha \rho \sigma}, \;
\mu^{-2}R_{\rho \nu}T^{\rho \nu}, \;
\mu^{-4}T_{\rho \nu}T^{\rho \nu}
$$
as well, where $T^{\rho \nu}$ is the Yang--Mills stress-energy
density in curved spacetime, and $\mu$ is the Planck mass. The
presence of the last $3$ terms means that, although the Yang--Mills
coupling constant gets renormalized, the finite part of the
effective action now depends on the auxiliary mass in a way that
cannot be absorbed into a running coupling constant. Each choice of
auxiliary mass corresponds to a different theory \cite{dewi03}.
\vskip 0.3cm
\noindent
(iv) The Yang--Mills field, in turn, spoils a basic property of
pure gravity based on Einstein's general relativity. Indeed,
although many remarkable cancellations occur in the computation,
the presence of the Yang--Mills field destroys \cite{dewi03}
the one-loop finiteness of pure gravity \cite{thooft74}.

Section II outlines the method proposed in Ref. \cite{mdc13}
to study the effects of curvature on the size of the Gribov region.
Sections III and IV study the effects of Ricci and Riemann tensor.
The challenge of evaluating gluon and ghost propagators in curved
spacetime is analyzed in Sec. V, while our results are interpreted 
in Sec. VI.

\section{Effects of the curvature on the size of the Gribov region}

Consider a point $p$ on a given spacetime $(M,g)$, and choose Riemann
normal coordinates in a neighbourhood of $p$. Such a coordinate system
is built as follows. For each $X \in U \subset T_{p}(M)$, consider
the affinely parametrized geodesic $\gamma_{X}$ starting from $p$
with initial velocity $X$. By definition, the exponential map is such
that 
$$
{\rm e}^{X}: \; X \in U \rightarrow q=\gamma_{X}(1).
$$
The points $q$ form a neighbourhood $I$ of $p$. If $U$ is sufficiently
small, the exponential is invertible and one can use coordinates
of the vector $X$ in $T_{p}(M)$ to identify the point $q$. In such a
coordinate system, if one takes the coordinate lines to be orthogonal
at $p$, the metric tensor $g_{\mu \nu}$ and Christoffel symbols
$\Gamma_{\mu \nu}^{\lambda}$ are given by the following approximate
formulae:
\begin{equation}
g_{\mu \nu}=\delta_{\mu \nu}-{1\over 3}R_{\mu \alpha \nu \beta}
X^{\alpha}X^{\beta}+{\rm O}(\| X \|^{3}),
\end{equation}
\begin{equation}
\Gamma_{\mu \nu}^{\lambda}=-{1\over 3}
\left(R_{\; \mu \nu \beta}^{\lambda}
+R_{\; \nu \mu \beta}^{\lambda}\right)X^{\beta}
+{\rm O}(\|X \|^{2}).
\end{equation}
Thus, in Riemannian geodesic coordinates, spacetime displays only a
tiny deviation from flatness. If the gravitational field, described by
the Riemann curvature tensor, is weak, one can use perturbation theory
to study the modifications introduced by a non-vanishing Riemann tensor.

The ghost operator for quantum Yang--Mills theory in curved spaces
reads as (omitting hereafter, for simplicity of notation, 
Lie-algebra indices)
\begin{equation}
FP(A)=-\nabla_{\mu}\nabla^{\mu}-[A_{\mu},\nabla^{\mu}].
\end{equation}
It acts on anticommuting scalar fields, and reads eventually, in
the above coordinates,
\begin{equation}
FP(A)\omega=-g^{\mu \nu}\left(\partial_{\mu}\partial_{\nu}
\omega-\Gamma_{\mu \nu}^{\lambda}\partial_{\lambda}\omega
\right)-[A_{\mu},\partial^{\mu}\omega].
\end{equation}
By virtue of the formula expressing Christoffel symbols in Riemann
normal coordinates, one finds
\begin{equation}
\delta^{\mu \nu}\Gamma_{\mu \nu}^{\lambda}
={2\over 3}R_{\alpha}^{\lambda}X^{\alpha},
\end{equation}
and hence
\begin{equation}
FP(A)\omega=\left(-\Box + {2\over 3}R_{\alpha}^{\lambda}X^{\alpha}
\partial_{\lambda}-{1\over 3}R_{\; \alpha \; \beta}^{\mu \; \nu}
X^{\alpha}X^{\beta}\partial_{\mu}\partial_{\nu}\right)\omega
-[A_{\mu},\partial^{\mu}\omega].
\end{equation}
The second and third term, involving Ricci and Riemann, respectively,
are corrections which account for the presence of the gravitational
field.

\section{Effect of the Ricci term}

Suppose now that $\omega$ is a real zero-mode of the flat ghost
operator. We can thus use perturbation theory to evaluate the shift
to the zero-energy level. The Ricci contribution, denoted by
$\varepsilon_{1}$, takes the form
\begin{eqnarray}
\varepsilon_{1} & \equiv & {2\over 3}{\int R_{\alpha}^{\lambda}X^{\alpha}
{\rm Tr}(\omega \partial_{\lambda} \omega) \over 
\int {\rm Tr} (\omega^{2})}
={2\over 3}R_{\alpha}^{\lambda}{\int X^{\alpha} {1\over 2}
\partial_{\lambda}{\rm Tr}(\omega^{2}) \over 
\int {\rm Tr} (\omega^{2})} \nonumber \\
&=& -{1\over 3}R_{\alpha}^{\lambda}
{\int (\partial_{\lambda}X^{\alpha}){\rm Tr}(\omega^{2}) \over
\int {\rm Tr}(\omega^{2})}
=-{1\over 3}R_{\alpha}^{\lambda}\delta_{\lambda}^{\alpha}
=-{R\over 3},
\end{eqnarray}
where, by choosing Dirichlet boundary conditions for the ghost field,
we have been able to set to zero the boundary term after 
integration by parts (see the work in Ref. \cite{avra99} for a
detailed discussion of these ghost boundary conditions in quantum
Yang--Mills theory).

\section{Contribution of the Riemann term} 

No conclusion can be reached without a proper treatment of the
Riemann term since, as will be shown below, it also involves a term
linear in $X$. Indeed, the Riemann tensor contributes through
\begin{equation}
\varepsilon_{2} \equiv -{1\over 3}R_{\; \alpha \; \beta}^{\mu \; \nu}
{\int X^{\alpha}X^{\beta}{\rm Tr}(\omega \partial_{\mu}
\partial_{\nu} \omega) \over \int {\rm Tr}(\omega^{2})}.
\end{equation}

At this stage, we first use the identity
\begin{equation}
{\rm Tr}(\omega \partial_{\mu}\partial_{\nu}\omega)
={1\over 2}\partial_{\mu}\partial_{\nu}{\rm Tr}(\omega^{2})
-{\rm Tr}\Bigr((\partial_{\mu}\omega)(\partial_{\nu}\omega)\Bigr)
\end{equation}
to re-express $\varepsilon_{2}$ in the form
\begin{equation}
\varepsilon_{2}=-{1\over 6}R_{\; \alpha \; \beta}^{\mu \; \nu}
{\int X^{\alpha}X^{\beta}\partial_{\mu}\partial_{\nu}
{\rm Tr}(\omega^{2}) \over \int {\rm Tr}(\omega^{2})}
+{1\over 3}R_{\; \alpha \; \beta}^{\mu \; \nu}
{\int X^{\alpha}X^{\beta}{\rm Tr}((\partial_{\mu}\omega)
(\partial_{\nu}\omega)) \over
\int {\rm Tr}(\omega^{2})}.
\end{equation}

As a second step, we exploit the Leibniz rule, that provides
\begin{equation}
\partial_{\mu}\Bigr(X^{\alpha}X^{\beta}\partial_{\nu}{\rm Tr}
(\omega^{2})\Bigr)=\partial_{\mu}(X^{\alpha}X^{\beta})
\partial_{\nu}{\rm Tr}(\omega^{2})
+X^{\alpha}X^{\beta}\partial_{\mu}\partial_{\nu}
{\rm Tr}(\omega^{2}),
\end{equation}
to integrate by parts. Hence we find 
\begin{eqnarray}
\; & \; & \int X^{\alpha}X^{\beta}\partial_{\mu}\partial_{\nu}
{\rm Tr}(\omega^{2})=\int \partial_{\mu}\Bigr(X^{\alpha}X^{\beta}
\partial_{\nu}{\rm Tr}(\omega^{2})\Bigr) 
- \int \Bigr(\delta_{\mu}^{\alpha}X^{\beta}
+\delta_{\mu}^{\beta}X^{\alpha}\Bigr)\partial_{\nu}
{\rm Tr}(\omega^{2}) \nonumber \\
&=& -\delta_{\mu}^{\alpha} \int X^{\beta} \partial_{\nu}
{\rm Tr}(\omega^{2})
-\delta_{\mu}^{\beta} \int X^{\alpha} \partial_{\nu}{\rm Tr}
(\omega^{2}),
\end{eqnarray}
where we have again exploited Dirichlet boundary conditions for the
ghost field, bearing in mind that
$\partial_{\nu}{\rm Tr}(\omega^{2})
=2{\rm Tr}\Bigr(\omega \partial_{\nu}\omega \Bigr)$.

As a third step, we use again the Leibniz rule, i.e.
\begin{equation}
\partial_{\nu}\Bigr(X^{\alpha}{\rm Tr}(\omega^{2})\Bigr)
=\delta_{\nu}^{\alpha}{\rm Tr}(\omega^{2})
+X^{\alpha}\partial_{\nu}{\rm Tr}(\omega^{2}),
\end{equation}
and the same with $X^{\alpha}$ replaced by $X^{\beta}$, to find, by
virtue of Dirichlet boundary conditions for $\omega$,
\begin{equation}
\int X^{\alpha}X^{\beta}\partial_{\mu}\partial_{\nu}
{\rm Tr}(\omega^{2})=\Bigr(\delta_{\mu}^{\alpha}\delta_{\nu}^{\beta}
+\delta_{\nu}^{\alpha}\delta_{\mu}^{\beta}\Bigr)
\int {\rm Tr}(\omega^{2}).
\end{equation}
The first term on the right-hand side in our formula for 
$\varepsilon_{2}$ is therefore
\begin{equation}
{\widetilde \varepsilon}_{2}=-{1\over 6}
\Bigr(R_{\; \mu \; \beta}^{\mu \; \beta}
+R_{\; \nu \; \mu}^{\mu \; \nu}\Bigr)
={R \over 6},
\end{equation}
which implies that
\begin{equation}
\varepsilon_{2}={R \over 6}+{1\over 3}
{\int \gamma^{\mu \nu}{\rm Tr}(\omega_{,\mu}\omega_{,\nu})
\over \int {\rm Tr}(\omega^{2})},
\end{equation}
having defined
\begin{equation}
\gamma^{\mu \nu} \equiv R_{\; \alpha \; \beta}^{\mu \; \nu}
X^{\alpha}X^{\beta}.
\end{equation}
For example, in the so-called Euclidean de Sitter space, on 
denoting by $K$ a constant, $\gamma^{\mu \nu}$ reads as
\begin{equation}
\gamma^{\mu \nu}=K \Bigr(g^{\mu \nu}g_{\alpha \beta}
-\delta_{\beta}^{\mu} \delta_{\alpha}^{\nu}\Bigr)X^{\alpha}X^{\beta}
=K \Bigr(g^{\mu \nu} \|X \|^{2}-X^{\mu}X^{\nu}\Bigr),
\end{equation}
where $\| X \|^{2}=g(X,X)=X_{\alpha}X^{\alpha} >0$. Our 
$\gamma^{\mu \nu}$ acts as $g(X,X)g^{\mu \nu}$ on the hyperplane
orthogonal to $X$, while it vanishes on the sub-space generated
by $X$. Hence the sign of $\gamma^{\mu \nu}$ is ruled by the constant
$K$, which is positive.

\section{The challenge of gluon and ghost propagators}

When Yang-Mills theory is studied in flat space, it is rather important
to evaluate the gluon and ghost propagators, since their behavior in
the infrared depends crucially on the Gribov mass parameter, which is
determined through the so-called gap equation. More precisely, 
investigations of lattice gauge theory on very large volumes 
\cite{Cucchieri} have found an infrared finite gluon propagator and
a ghost propagator which is no longer enhanced
in the infrared. The work in Ref. \cite{Dudal} has exploited a refinement
of the Gribov-Zwanziger method to obtain analytical results in
agreement with these lattice data. Moreover, the work in Ref.
\cite{Fischer} has obtained, in various truncations of Dyson-Schwinger
equations and functional renormalization-group equations, a 
one-parameter family of solutions for the ghost and gluon dressing
functions of Landau gauge Yang-Mills theory, each member of the
one-parameter family being confining. 
In a general curved spacetime, however, no momentum space representation
is available in the first place, since the homogeneity required for its
existence is lacking, and the local momentum space formalism built
in Ref. \cite{Parker} is only appropriate for studying ultraviolet
divergences. Thus, one needs a radical departure from the 
calculational techniques available in flat space. 

For this purpose, we have carefully considered the 
Gusynin \cite{Gusynin} technique, which relies
in turn on the Widom \cite{Widom} formalism. 
Following Gusynin, one can express the matrix elements of 
the resolvent of a positive elliptic operator $H$ by means of the formula
\begin{equation}
\label{Widom}
G(x,x',\lambda)\equiv\langle x|\frac{1}{(H-\lambda {\rm I})}|x'\rangle
=\int\frac{d^{n}k}{(2\pi)^{n}\sqrt{g(x')}}e^{il(x,x',k)}\sigma(x,x',k;\lambda).
\end{equation}
Here $l(x,x',k)$ is a biscalar under general coordinate 
transformations and constitutes a generalization of the phase 
$k_{\mu}(x-x')^{\mu}$ used in the flat case. This expression for the 
resolvent is manifestly covariant.
The generalization of the linearity property of $l(x,x',k)$ valid in 
the flat case is obtained by requiring that symmetric combinations of 
covariant derivatives should vanish in the coincidence limit
\begin{equation}
\label{cov der symm}
\nabla_{(\mu_1}\nabla_{\mu_2}\dots\nabla_{\mu_{m})}l|_{x=x'}
\equiv\left[\nabla_{(\mu_1}\nabla_{\mu_2}\dots\nabla_{\mu_{m})}l\right]=0, 
\hspace{1em}\mbox{$m\neq 1$}.
\end{equation}
along with
\begin{equation}
\label{prima cov der}
[\nabla_{\mu}l]=k_{\mu}.
\end{equation}
The square bracket denotes the coincidence limit and symmetrization must 
be understood over indices enclosed by the round brackets. These conditions 
are sufficient to determine $l(x,x',k)$ in a neighborhood of the point 
$x'$. Indeed, the commutator of covariant derivatives acts on tensors as 
follows:
\begin{align}
[\nabla_{\mu},\nabla_{\nu}]f^{\nu_{1}\dots\nu_{k}}_{\mu_{1}\dots\mu_{n}}
=&R_{\mu\nu\lambda}^{\quad\;\nu_{i}}f^{\nu_{1}\dots\nu_{i-1}\lambda\nu_{i+1}
\dots\nu_{k}}_{\mu_{1}\dots\mu_{n}}-R_{\mu\nu\mu_{i}}^{\quad\;\lambda}
f^{\nu_{1}\dots\nu_{k}}_{\mu_{1}\dots\mu_{i-1}\lambda\mu_{i+1}
\dots\mu_{n}} \\
+&T^{\lambda}_{\mu\nu}\nabla_{\lambda}f^{\nu_{1}\dots\nu_{k}}_{\mu_{1}
\dots\mu_{n}}+W_{\mu\nu}f^{\nu_{1}\dots\nu_{k}}_{\mu_{1}\dots\mu_{n}}.
\end{align}
Using this formula and (\ref{cov der symm}), (\ref{prima cov der}) one 
can find the covariant derivatives of $l$ in the coincidence limit. 
The resolvent kernel $G(x,x',\lambda)$ is a solution of the equation
\begin{equation}
(H(x,\nabla_{\mu})-\lambda)G(x,x',\lambda)=\frac{1}{\sqrt{g}}\delta(x-x'),
\end{equation}
subject to the boundary conditions which define the domain of the operator 
$H$. By inserting into this equation the integral formula for the resolvent 
kernel one gets the equation
\begin{equation}
\label{equazione per sigma}
(H(x,\nabla_{\mu}+i\nabla_{\mu}l)-\lambda)\sigma(x,x',k;\lambda)=I(x,x').
\end{equation}
The function $I(x,x')$ is a biscalar and is defined by conditions analogous 
to those satisfied by $l(x,x',k)$
\begin{align}
[I]=&E,\\
\left[\nabla_{(\mu_1}\nabla_{\mu_2}\dots\nabla_{\mu_{m})}I\right]=&0,
\end{align}
where $E$ is the unit matrix.

One then introduces an auxiliary parameter $\varepsilon$, which will be 
set to $1$ at the end of the calculations, and  
$\sigma(x,x',k;\lambda)$ and $H(x,\nabla_{\mu}+i\nabla_{\mu}l)$ 
are expanded by following the rules 
$l\to l/\varepsilon$, $\lambda\to\lambda/\varepsilon^{2r}$, i.e.
\begin{align}
\sigma_{\varepsilon}(x,x',k;\lambda)=&\sum_{m=0}^{\infty}\varepsilon^{2r+m}
\sigma_{m}(x,x',k;\lambda), \\
H(x,\nabla_{\mu}+i\nabla_{\mu}l/\varepsilon)=&\sum_{m=0}^{2r}
\varepsilon^{-2r+m}A_{m}(x,\nabla_{\mu},\nabla_{\mu}l).
\end{align}
Substituting these expansions into Eq. (\ref{equazione per sigma}) 
and collecting terms of the same order in $\varepsilon$ one gets a system 
of equations for the coefficients $\sigma_{m}$ 
which can be solved recursively.

The diagonal matrix elements of the heat-kernel are 
then given by the relation
\begin{equation}
\langle x | e^{-tH}| x \rangle
=\sum_{m=0}^{\infty}\int \frac{d^{n}k}{(2\pi)^{n}\sqrt{g}}
\int_{C}\frac{id\lambda}{2\pi}e^{-t\lambda}[\sigma_{m}](x,k,\lambda).
\end{equation}

One finds from the recursion relations satisfied by the coefficients 
$\sigma_{m}$, that these coefficients are generalized homogeneous functions 
in the variables $(k,\lambda)$
\begin{equation}
[\sigma_{m}](x,tk,t^{2r}\lambda)]=t^{-(m+2r)}[\sigma_{m}](x,k,\lambda)].
\end{equation}
Hence it follows that the heat-kernel expansion coefficients are obtained 
from those of the Laplace transform of the resolvent kernel
\begin{equation}
E_{m}(x|H)=\int \frac{d^{n}k}{(2\pi)^{n}\sqrt{g}}\int_{C}
\frac{id\lambda}{2\pi}e^{-t\lambda}[\sigma_{m}](x,k,\lambda).
\end{equation}
Conversely, the resolvent kernel may be obtained from the heat
kernel $K(x,x',t)$ by the Laplace transform
\begin{equation}
G(x,x',\lambda)=\int_{0}^{\infty}e^{t \lambda}K(x,x',t)dt,
\end{equation}
and the Green function ${\cal G}(x,x')$ is equal to $G(x,x',\lambda=0)$.

The advantage of this approach is that it gives an algorithm to calculate 
the coefficients $E_{m}(x|H)$ and it can be generalized to the case of 
non-minimal operators. Non-minimal second-order operators are indeed 
a very interesting class of operators, whose general form is
\begin{equation}
H^{\mu\nu}=-g^{\mu\nu}\Box+a\nabla^{\mu}\nabla^{\nu}+X^{\mu\nu}.
\end{equation}
Here $\nabla^{\mu}$ is the covariant derivative, including both the 
Levi-Civita connection and the gauge connection. The tensor $X^{\mu\nu}$ is 
a matrix in the internal indices. The parameter $a$ may assume all real 
values, in particular for $a=0$ the operator reduces to a minimal one. The 
gauge-field operator for Yang-Mills' theory falls indeed in this class
\begin{equation}
\label{H operator YM}
H^{\mu\nu}_{YM}=-g^{\mu\nu}\Box+\left(1-\frac{1}{\alpha}\right)
\nabla^{\mu}\nabla^{\nu}+R^{\mu\nu}.
\end{equation}
This can also be expressed as an operator acting on $1$-forms
\begin{equation}
H(\alpha)=\delta d+\frac{1}{\alpha} d\delta.
\end{equation}
In the case $\alpha=1$ it reduces to the Laplace-Beltrami operator, whose 
action on $1$-forms $\varphi_{\nu}dx^{\nu}$ is given by a 
Bochner-Lichnerowicz formula
\begin{equation}
\left(\left(\delta d+\frac{1}{\alpha} d\delta\right)\varphi\right)_{\mu}
=(-\delta_{\mu}^{\;\nu}\Box+R_{\mu}^{\;\nu})\varphi_{\nu}.
\end{equation}
We also recall that Endo \cite{Endo} obtained a 
formula which makes it possible to express the integrated 
heat-kernel for a generic value of the parameter $\alpha$ in 
terms of that relative to the minimal case $\alpha=1$, i.e.
\begin{equation}
K^{(\alpha)}_{\mu\nu'}(\tau)=K^{(1)}_{\mu\nu'}(\tau)
+{\rm i}\int_{\tau}^{\tau/\alpha}dy\;\nabla_{\mu}
\nabla^{\lambda}K^{(1)}_{\lambda\nu'}(y).
\end{equation}

Unfortunately, the heat-kernel expansion corresponding to our 
gauge-field operator does not exist in the singular case 
$\alpha \to 0$, i.e. the Landau choice of gauge parameter. 
In fact in this limit only some heat-kernel coefficients 
are finite, while the others diverge. 
A possible way out might be
to remove the divergent parts of such coefficients. More precisely,
the work in Ref. \cite{Faizal} has evaluated the ghost propagator
for Yang-Mills in de Sitter space (see Eqs. (2.5) and (2.6) therein). 
The authors of Ref. \cite{Faizal} argue that, since ghost fields occur
only in internal loops and couple to the gauge field through a
derivative coupling, the divergent term in the ghost propagator does
not contribute to the calculation of $n$-point functions of gauge fields.
Hence they propose that one should use the effective ghost propagator
obtained by subtracting the divergent contribution. We are currently
trying to understand whether such a subtraction procedure can be 
advocated for both gluon and ghost propagators also when  
Yang-Mills theory is studied in a generic curved spacetime in the
presence of a Gribov horizon.   

\section{Interpretation and concluding remarks}

Within our scheme, the full shift to the zero-energy level reads as
\begin{equation}
\varepsilon_{1}+\varepsilon_{2}=-{R \over 6}
+{1\over 3}{\int \gamma^{\mu \nu}{\rm Tr}(\omega_{,\mu}
\omega_{,\nu}) \over \int {\rm Tr}(\omega^{2})}.
\end{equation}
If it were just for the $-{R \over 6}$ term, we might argue
as follows \cite{mdc13}: if the scalar curvature $R$ is positive,
the Gribov horizon moves inward, hence it is reached at a higher
energy and the gauge-field propagator should be more suppressed.
By contrast, if $R$ is negative, the horizon moves outward, and the
energy such that field fluctuations reach the horizon should be
lower. In other words, the gauge-field propagator should be less
suppressed in the infrared if $R<0$. For these conclusions to remain
qualitatively the same, we should study the conditions
$\varepsilon_{1}+\varepsilon_{2}<0$ and
$\varepsilon_{1}+\varepsilon_{2}>0$, respectively. The former is
fulfilled provided that
\begin{equation}
2{\int \gamma^{\mu \nu}{\rm Tr}(\omega_{,\mu}\omega_{,\nu}) \over
\int {\rm Tr}(\omega^{2})} < R,
\end{equation}
while the latter is satisfied in the opposite case, i.e. if
\begin{equation}
2{\int \gamma^{\mu \nu}{\rm Tr}(\omega_{,\mu}\omega_{,\nu}) \over
\int {\rm Tr}(\omega^{2})} > R.
\end{equation}
For each choice of curved Riemannian background, one has to check
which of the two conditions above is satisfied. It should be
stressed that it is hard to obtain an estimate of the integral on
the left-hand side of (6.2) and (6.3), because the zero-mode 
$\omega$, and hence the integral itself, depends on the gauge 
connection at the Gribov horizon. Such a difficulty becomes clearer
if one bears in mind that the Gribov horizon is not precisely 
localizable, not even in the flat case (where it is known that, in
a first approximation, it is an ellipsoid \cite{mdc13}). 
Nevertheless, we hope that the scheme here proposed, 
with the explicit computational recipe provided, will
lead to further progress on the understanding of the Gribov phenomenon
in curved spaces. It would be also interesting to study the extension
to curved spacetime of the scheme proposed in Ref. \cite{Pere13}
for the elimination of infinitesimal Gribov ambiguities in nonAbelian
gauge theories. 

\acknowledgments 
G. Esposito is grateful to the Dipartimento di Fisica of
Federico II University, Naples, for hospitality and support. 
H. Ghorbani would like to thank INFN Sezione di Napoli for hospitality
and financial support to his visit.


\begin{thebibliography}{}

\bibitem{gribov78}
V. N. Gribov, Nucl. Phys. B {\bf 139}, 1 (1978).
\bibitem{halp78}
M. B. Halpern and J. Koplik, Nucl. Phys. B {\bf 132}, 239 (1978).
\bibitem{espo04}
G. Esposito, D. N. Pelliccia, and F. Zaccaria, Int. J. Geom. Methods
Mod. Phys. {\bf 1}, 423 (2004).
\bibitem{mdc13}
M. de Cesare, arXiv:1308.5344 [hep-th], 
to appear in Int. J. Geom. Methods Mod. Phys. {\bf 11} (2014).
\bibitem{canfo10}
F. Canfora, A. Giacomini, and J. Oliva, Phys. Rev. D {\bf 82},
045014 (2010).
\bibitem{canfo11}
F. Canfora, A. Giacomini, and J. Oliva, Phys. Rev. D {\bf 84},
105019 (2011).
\bibitem{dewi03}
B. S. DeWitt, {\it The Global Approach to Quantum Field Theory},
International Series of Monographs on Physics (Oxford University
Press, Oxford, 2003), Vol. 114.
\bibitem{thooft74}
G. 't Hooft and M. Veltman, Ann. Inst. Henri Poincar\'e {\bf 20},
69 (1974).
\bibitem{avra99}
I. G. Avramidi and G. Esposito, Commun. Math. Phys. {\bf 200},
495 (1999).
\bibitem{Cucchieri}
A. Cucchieri and T. Mendes, Phys. Rev. Lett. {\bf 100}, 241601 (2008);
Phys. Rev. D {\bf 78}, 094503 (2008).
\bibitem{Dudal}
D. Dudal, J. A. Gracey, S. P. Sorella, N. Vandersickel, and
H. Verschelde, Phys. Rev. D {\bf 78}, 065047 (2008).
\bibitem{Fischer}
C. Fischer, A. Maas, and J. M. Pawlowski, Ann. Phys. {\bf 324},
2408 (2009).
\bibitem{Parker}
T. S. Bunch and L. Parker, Phys. Rev. D {\bf 20}, 2499 (1979).
\bibitem{Gusynin}
V. P. Gusynin, Phys. Lett. B {\bf 225}, 233 (1989).
\bibitem{Widom}
H. Widom, Bull. Sci. Math. {\bf 104}, 19 (1980).
\bibitem{Endo}
R. Endo, Prog. Theor. Phys. {\bf 71}, 1366 (1984).
\bibitem{Faizal}
M. Faizal and A. Higuchi, Phys. Rev. D {\bf 78}, 067502 (2008).
\bibitem{Pere13}
A. D. Pereira and R. F. Sobreiro, arXiv:1308.4159 [hep-th].
\end{thebibliography}
\end{document}